\documentclass[twocolumn]{aastex62}
\usepackage{amssymb}
\usepackage{array,multirow}
\usepackage{comment}
\usepackage{enumerate}
\usepackage{bm}
\usepackage{threeparttable}

\newcommand{\Porb}{\ifmmode {P_{\rm orb}}\else${P_{\rm orb}}$\fi}

\newcommand{\Msun}{\ifmmode {{M_\odot}}\else{$M_\odot$}\fi}
\newcommand{\Rsun}{\ifmmode {{R_\odot}}\else{$R_\odot$}\fi}
\newcommand{\Mtot}{\ifmmode {{M_{\rm tot}}}\else{$M_{\rm tot}$}\fi}
\newcommand{\RV}{\ifmmode {{\rm RV}}\else RV \fi}
\newcommand{\bigG}{\ifmmode {\mathcal{G}}\else${\mathcal{G}}$\fi}

\submitjournal{ApJ}

\shorttitle{Gaia DR3 Binaries}

\begin{document}

\title{{\bf A Sample of Neutron Star and Black Hole Binaries Detected through Gaia DR3 Astrometry}}

\author[0000-0001-5261-3923]{Jeff J. Andrews}
\affiliation{Center for Interdisciplinary Exploration and Research in Astrophysics (CIERA), 
1800 Sherman Ave., 
Evanston, IL, 60201, USA}
\affiliation{Department of Physics, University of Florida, 2001 Museum Rd., Gainesville, FL 32611}
\email{jeffrey.andrews@northwestern.edu}

\author[0000-0002-5748-4558]{Kirsty Taggart}
\affiliation{Department of Astronomy and Astrophysics, University of California, Santa Cruz, CA 95064, USA}

\author[0000-0002-2445-5275]{Ryan~J.~Foley}
\affiliation{Department of Astronomy and Astrophysics, University of California, Santa Cruz, CA 95064, USA}

\begin{abstract}
With its exquisite astrometric precision, the latest Gaia data release includes $\sim${}$10^5$ astrometric binaries, each of which have measured orbital periods, eccentricities, and the Thiele-Innes orbital parameters. Using these and an estimate of the luminous stars' masses, we derive the companion stars' masses, from which we identify a sample of 24 binaries in long period orbits ($P_{\rm orb}~\sim~{\rm yrs}$) with a high probability of hosting a massive ($>$1.4~$\Msun$), dark companion: a neutron star (NS) or black hole (BH). The luminous stars in these binaries tend to be F-, G-, and K-dwarfs with the notable exception of one hot subdwarf. Follow-up spectroscopy of eight of these stars shows no evidence for contamination by white dwarfs or other luminous stars. The dark companions in these binaries span a mass range of 1.35--2.7 $\Msun$ and therefore likely includes both NSs and BHs without a significant mass gap in between. 
Furthermore, the masses of several of these objects are $\simeq$1.7 $\Msun$, similar to the mass of at least one of the merging compact objects in GW190425. Given that these orbits are too wide for significant mass accretion to have occurred, this sample implies that some NSs are born heavy ($\gtrsim$1.5~\Msun). Additionally, the low orbital velocities ($\lesssim$20~km s$^{-1}$) of these binaries requires that at least some heavy NSs receive low natal kicks, otherwise they would have been disrupted during core collapse. Although none will become gravitational wave sources within a Hubble time, these systems will be exceptionally useful for testing binary evolution theory.
\end{abstract}


\section{Introduction}
\label{S:intro}

Neutron stars (NSs) and black holes (BHs) form the engines for some of the most exotic astrophysical environments, allowing for tests of general relativity \citep{eht}, nuclear matter at high densities \citep{akmal1998}, and extreme gas physics \citep{woosley1993}. The best studied examples of these objects are found when they orbit another star \citep[e.g.,][]{hulse1975}, as companion stars can donate mass to an accreting NS or BH \citep{tauris2006} and can provide dynamically derived mass measurements \citep{demorest2010}. Therefore, it is of great astrophysical interest to identify more of these systems, across a broad range of environments.

With the detection of GW150914, the first binary black hole merger through gravitational waves, a new window into BH science opened up \citep{GW150914}. The latest catalog of gravitational wave events contains nearly a hundred separate detections, each with their own measured masses and spin constraints \citep{gwtc3}. Analysis of this population has led to a more complete understanding of merging BHs in the Universe \citep{gwtc3-population, van_son2022}. Despite the size of this catalog, significant questions remain, so that even the dominant formation scenario is uncertain \citep{wong2021, zevin2021}. Electromagnetic observations provide an alternative tool for learning about compact object binaries.

The depth of their potential wells leads to accreting BHs emitting copious X-rays \citep{shakura1973}, and indeed nearly every known stellar mass BH has been found through X-ray observations \citep{remillard2006}. While accreting NSs also emit X-rays, non-accreting NSs can themselves emit pulsed radio emission \citep{ruderman1975}, which has led to a rich landscape of NS binaries ranging from the so-called spider pulsars \citep[with companions with masses of a few $10^{-2}$ to a few $10^{-1}$ \Msun;][]{roberts2013} to double neutron stars \citep[which host other NSs as their companions;][]{tauris2017}. Despite the success of X-ray and radio observations at identifying BH and NS binaries, observers have explored other methods to identify these binaries.

Starting with pioneering studies by \citet{guseinov1966} and \citet{Trimble1969}, it was realized that stars exhibiting large radial velocity variations could indicate the presence of a massive, dark companion: a NS or BH. Compact objects accreting at a sufficiently low rate are unlikely to form a disk and therefore be very radiatively inefficient, emitting few X-rays \citep{hirai2021}. Several groups have begun to search time-series radial velocity catalogs for stars showing extreme variations, and a handful of detections have recently been claimed \citep{liu2019, thompson2019, rivinius2020, jayasinghe2021, lennon2021, jayasinghe2022, saracino2022}. However, upon further analysis most of these have been shown to be incorrect classifications \citep{abdul-masih2020, elbadry2020a, eldridge2020, irrgang2020, bodensteiner2020, shenar2020, elbadry2021a, elbadry2022a, elbadry2022b, elbadry2022c, stevance2022}. The only systems still suspected to host candidate BHs are 2MASS J05215658$+$4359220 \citep{thompson2019} and the BH binaries in the globular cluster NGC 3201 \citep{Giesers2018, giesers2019}. The BHs in NGC 3201 are a remarkable confirmation of model predictions of stellar dynamics in dense environments \citep{rodriguez2016} while isolated systems like 2MASS J05215658$+$4359220 provide critical tests of binary evolution theory \citep{breivik2019}.

While useful, radial velocity surveys tend to be an expensive method for searching for compact object binaries. Roughly simultaneously, \citet{Breivik2017} and \citet{Mashian2017} realized that compact object binaries could be identified astrometrically with the Gaia space telescope, by tracking the orbital motion of a luminous star as a NS or BH pulls it around \citep[see also][]{Barstow2014, Yalinewich2018, Yamaguchi2018}. The latest predictions include state-of-the-art binary population synthesis codes combined with a Milky Way model that includes dust extinction and an accurate model for Gaia's astrometric sensitivity \citep{andrews2019, chawla2021}. 

The latest Gaia third data release includes 34 months of data for $>${}$10^9$ stars \citep{gaia_EDR3}, from which the Gaia team have identified $>${}$10^5$ astrometric binaries \citep{GaiaDR3_binaries, halbwachs2022, holl2022}. Since astrometry allows for a measurement of the binary's orientation in space, including its inclination angle, the component masses can be better calculated than binaries measured with radial velocity alone \citep{andrews2019}. Therefore, we search the Gaia astrometric catalog for binaries that are likely to host NS or BH companions. 

In Section~\ref{sec:sample_selection} we describe the process by which we winnow down the sample of $\sim10^5$ astrometric binaries into our sample of massive, compact object binaries. We provide follow-up spectroscopy for a subset of our sample in Section~\ref{sec:observations}. We discuss our sample and provide some caveats in Section~\ref{sec:discussion}, and we conclude in Section~\ref{sec:conclusions}.

\section{Sample Selection}
\label{sec:sample_selection}

To select our sample, we start with the catalog of 134,598 binaries in the DR3 catalog of non-single stars that are detected using astrometry only \citep{GaiaDR3_binaries}. We focus on these binaries, avoiding spectroscopic and photometric binaries, as these have been discussed elsewhere \citep{elbadry2022d, gomel2022, mazeh2022}. For each of these binaries, we calculate the angular orbital separation of the luminous star, $a_0$, from the Thiele-Innes parameters, $A$, $B$, $F$, and $G$, provided in the non-single star catalog following the formula provided by \citet{halbwachs2022}:
\begin{eqnarray}
u &=& (A^2 + B^2 + F^2 + G^2)/2 \nonumber \\
v &=& AG - BF \nonumber \\
a_0 &=& \sqrt{ u + \sqrt{u^2-v^2} }.
\end{eqnarray}
Note that since each of the Thiele-Innes parameters has units of mas, $a_0$ is the angular orbital separation, also expressed in mas. We use the parallax, $\varpi$, to convert $a_0$ to a physical orbital separation and then use Kepler's third law and the observed orbital period, $P_{\rm orb}$, to calculate the astrometric mass function, $m_f$:
\begin{equation}
    m_f = 1 \left( \frac{a_0}{1\ {\rm mas}} \right)^3 \left(\frac{\varpi}{1\ {\rm mas}}\right)^{-3} \left( \frac{P_{\rm orb}}{1\ {\rm yr}} \right)^{-2}\ \Msun.
    \label{eq:mf_dark}
\end{equation}

Under the assumption that the companion is dark (which we test below) and therefore the photocenter of the system follows the observed star, the astrometric mass function is:
\begin{equation}
    m_f = \frac{M_2^3}{(M_1 + M_2)^2},
    \label{eq:mf}
\end{equation}
where the index 1 corresponds to the observed star (or at least the most luminous star, which we indicate as the primary in the system) and index 2 corresponds to its companion. Note that this mass function differs from the traditional mass function used for spectroscopic binaries by a factor $\sin^3 i$, where $i$ is the inclination angle of the orbit. 

\begin{figure}
    \begin{center}
    \includegraphics[width=1.0\columnwidth]{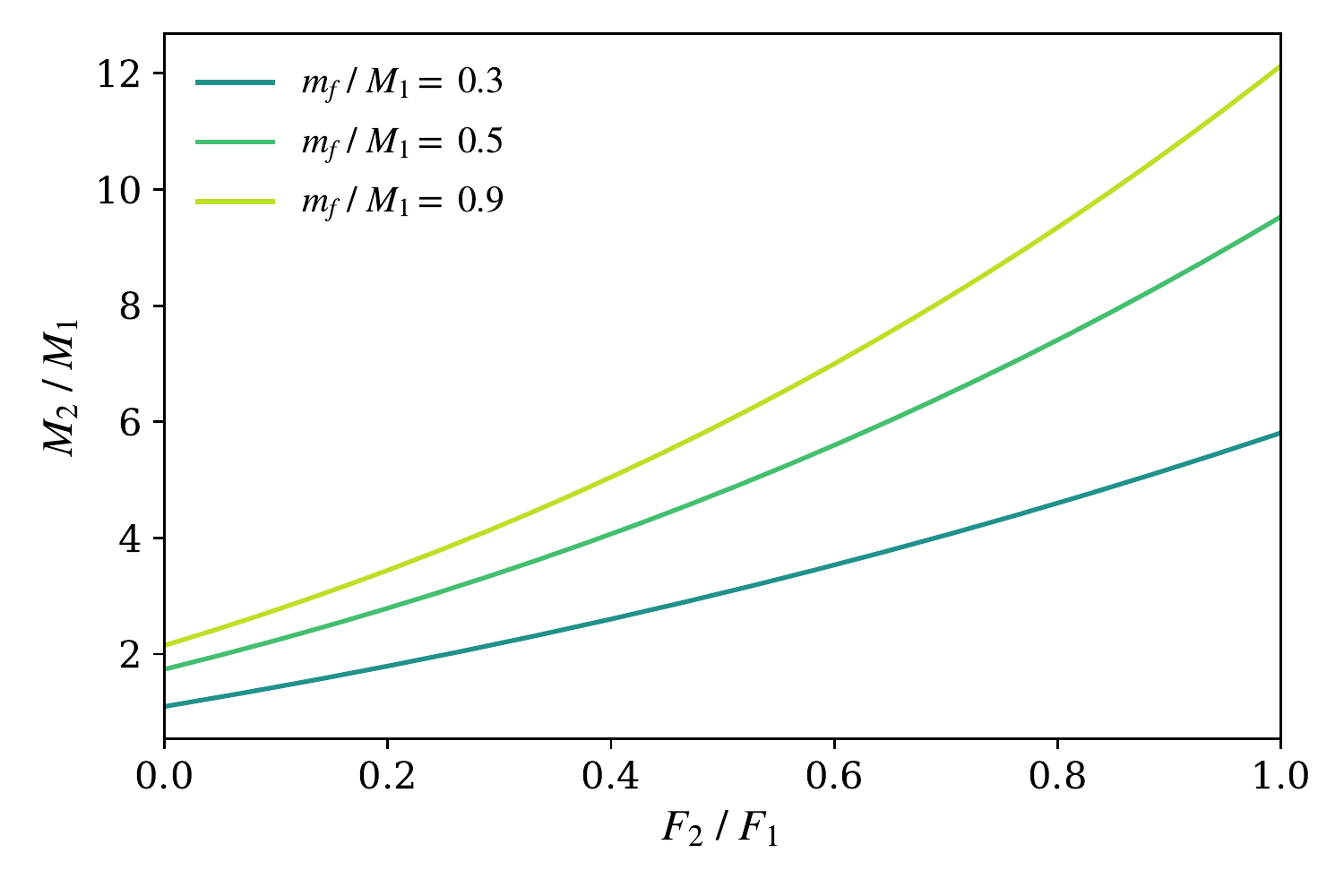}
    \caption{ Mass ratio as a function of the contribution of the companion's flux for three different mass functions as a function of the luminous star's mass. If our assumption of a dark companion is incorrect, then a larger companion mass is required. }
    \label{fig:mass_function}
    \end{center}
\end{figure}

Before proceeding, it is worth considering our assumption of a dark companion. Gaia observes the photocenter of the system, the photometric average position of the two stars. If the companion contributes to the luminosity of the system, we can express $m_f$ in its more general form \citep{halbwachs2022}:
\begin{equation}
    m_f = \frac{(F_1M_2 - F_2M_1)^3}{(F_1+F_2)^3 (M_1+M_2)^2}.
\end{equation}
Defining the flux ratio, $F=F_2/F_1$, and the mass ratio $q=M_2/M_1$, we can express $m_f$ as:
\begin{equation}
    \frac{m_f}{M_1} = \frac{(q-F)^3}{(1+F)^3(1+q)^2}. 
\end{equation}
For a given $m_f/M_1$, Figure~\ref{fig:mass_function} shows how the derived companion mass rapidly increases as its contribution to the overall flux increases; as the companion contributes more and more of the flux, the observed photocenter motion implies a larger and larger orbital separation--and therefore a more massive companion to match the observed orbital period. Our assumption of a dark companion is the most conservative one. We discuss the implications of this assumption further in Section~\ref{sec:discussion}.

Following the assumption of a dark companion, as a first cut, we calculate the mass function from Equation~\ref{eq:mf_dark} for all 134,598 binaries. Then, assuming the luminous star is 1~$\Msun$, we derive companion masses (this assumption is improved later in our analysis). We propagate uncertainties in the derived companion masses using $10^4$ Monte Carlo random draws of the full 12$\times$12 covariance matrices for each binary's astrometric solution. After selecting only binaries with a 95\% probability of having a companion more massive than 1.4~$\Msun$, we find 106 candidate binaries with NS or BH companions. This limiting mass is purposefully chosen to avoid contamination by massive white dwarfs. 

\begin{figure}
    \begin{center}
    \includegraphics[width=1.0\columnwidth]{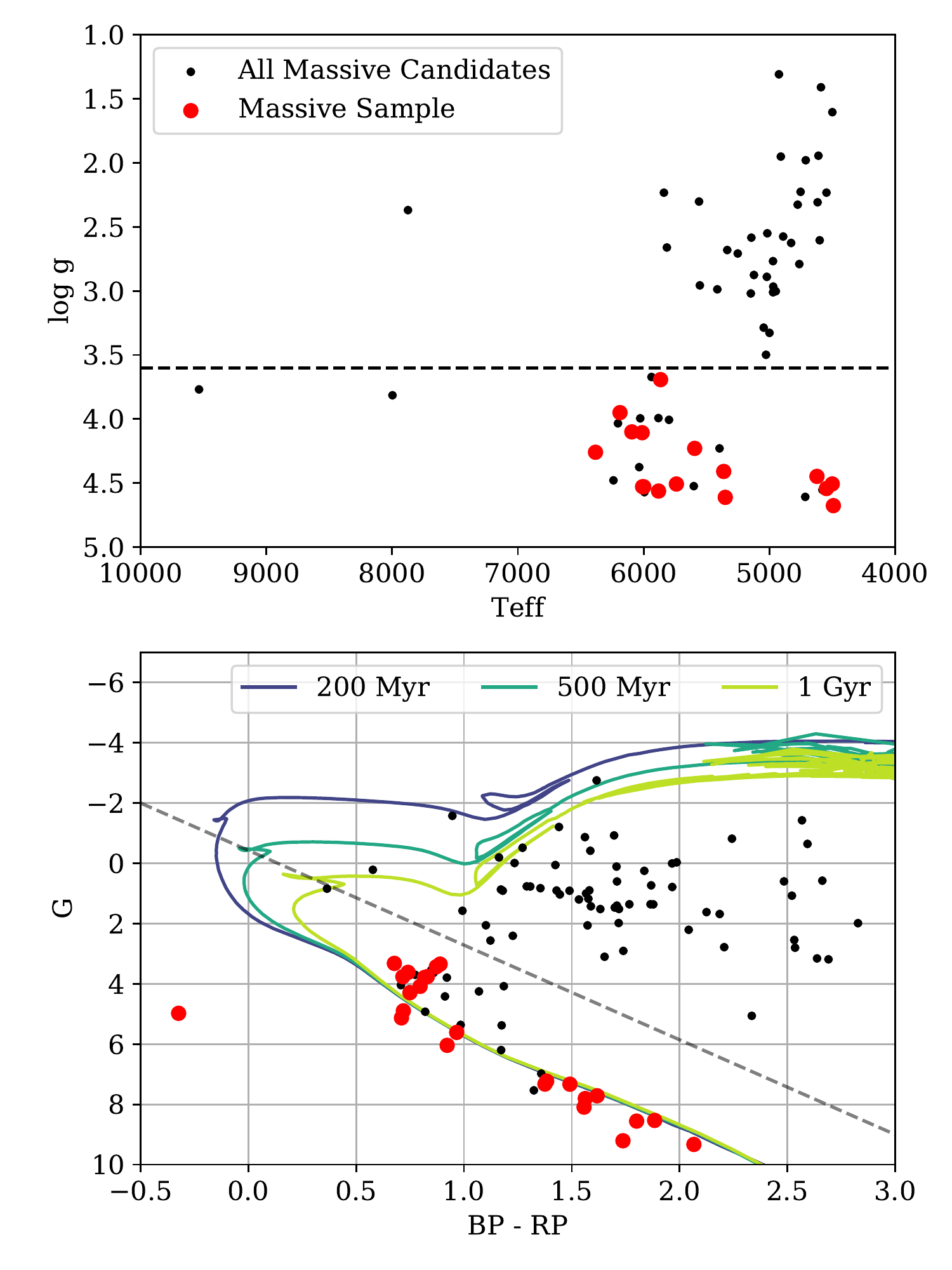}
    \caption{ The surface gravity and effective temperatures for the subset of the 106 candidate stars that are derived by Apsis (top panel). We indicate the stars in our final sample of 24 binaries as red points. Measurement uncertainties are typically $<$100 K and $<$0.1 for $T_{\rm eff}$ and log $g$, respectively, in the top panel. We additionally show (bottom panel) a color-magnitude diagram for the same two populations with PARSEC model isochrones overplotted \citep{bressan2012}. Black dashed lines in each panel indicate our restriction that we remove all giant stars. Typical Gaia photometric errors in the bottom panel are similar to or smaller than the plot markers. }
    \label{fig:dwarfs}
    \end{center}
\end{figure}

\begin{deluxetable*}{lccccccc}
 \tablecaption{ Catalog of Candidate Compact Object Binaries\tablenotemark{a} \label{tab:sample}}
 \tablehead{
    \colhead{Gaia DR3 Source ID} & 
    \colhead{$\varpi$} & 
    \colhead{Gaia $G$} & 
    \colhead{$P_{\rm orb}$} &
    \colhead{$e$} & 
    \colhead{$m_f$} & 
    \colhead{$M_1$\tablenotemark{b}} & 
    \colhead{$M_2$} \\
    \colhead{} &
    \colhead{(mas)} &
    \colhead{(mag)} &
    \colhead{(days)} &
    \colhead{} &
    \colhead{($M_{\odot}$)} &
    \colhead{($M_{\odot}$)} &
    \colhead{($M_{\odot}$)}
    }
\startdata
5681911574178198400 & 2.18$\pm$0.10 & 15.64 & 944$\pm$62 & 0.60$\pm$0.04 & 0.57$^{+0.09}_{-0.07}$ & 0.73$\pm$0.2 & 1.35$^{+0.19}_{-0.18}$ \\
3649963989549165440 & 1.37$\pm$0.10 & 14.30 & 893$\pm$120 & 0.36$\pm$0.28 & 0.79$^{+0.50}_{-0.23}$ & 0.47$\pm$0.2 & 1.41$^{+0.62}_{-0.34}$ \\
747174436620510976 & 2.58$\pm$0.06 & 13.99 & 999$\pm$53 & 0.71$\pm$0.04 & 0.56$^{+0.05}_{-0.04}$ & 0.85$\pm$0.2 & 1.43$^{+0.15}_{-0.16}$ \\
1581117310088807552 & 4.57$\pm$0.04 & 14.51 & 927$\pm$11 & 0.52$\pm$0.01 & 0.64$^{+0.03}_{-0.03}$ & 0.70$\pm$0.2 & 1.43$^{+0.16}_{-0.17}$ \\
1525829295599805184 & 4.04$\pm$0.11 & 16.18 & 328$\pm$2 & 0.37$\pm$0.04 & 0.69$^{+0.07}_{-0.07}$ & 0.64$\pm$0.2 & 1.44$^{+0.19}_{-0.20}$ \\
4271998639836225920 & 5.53$\pm$0.09 & 15.62 & 545$\pm$2 & 0.44$\pm$0.05 & 0.59$^{+0.07}_{-0.06}$ & 0.63 -- 1.00 & 1.44$^{+0.17}_{-0.17}$ \\
1695294922548180224 & 1.34$\pm$0.03 & 13.12 & 601$\pm$6 & 0.57$\pm$0.04 & 0.51$^{+0.04}_{-0.04}$ & 1.13$\pm$0.2 & 1.54$^{+0.14}_{-0.14}$ \\
1854241667792418304 & 4.41$\pm$0.04 & 14.87 & 1430$\pm$66 & 0.59$\pm$0.02 & 0.75$^{+0.03}_{-0.03}$ & 0.70$\pm$0.2 & 1.57$^{+0.17}_{-0.18}$ \\
1058875159778407808 & 0.90$\pm$0.06 & 14.52 & 836$\pm$59 & 0.42$\pm$0.08 & 0.69$^{+0.22}_{-0.12}$ & 0.63 -- 1.00 & 1.58$^{+0.33}_{-0.24}$ \\
1947292821452944896 & 1.81$\pm$0.09 & 15.94 & 1246$\pm$327 & 0.59$\pm$0.07 & 0.76$^{+0.26}_{-0.12}$ & 0.73$\pm$0.2 & 1.62$^{+0.35}_{-0.24}$ \\
2397135910639986304 & 2.04$\pm$0.05 & 13.35 & 916$\pm$38 & 0.56$\pm$0.06 & 0.58$^{+0.11}_{-0.09}$ & 1.10$\pm$0.2 & 1.62$^{+0.21}_{-0.19}$ \\
1144019690966028928 & 2.05$\pm$0.02 & 13.57 & 1402$\pm$123 & 0.38$\pm$0.04 & 0.59$^{+0.06}_{-0.05}$ & 1.08$\pm$0.2 & 1.63$^{+0.16}_{-0.15}$ \\
6593763230249162112 & 1.29$\pm$0.06 & 13.54 & 680$\pm$6 & 0.61$\pm$0.13 & 0.66$^{+0.22}_{-0.16}$ & 1.00$\pm$0.2 & 1.68$^{+0.34}_{-0.28}$ \\
5590962927271507712 & 1.95$\pm$0.08 & 15.88 & 818$\pm$10 & 0.72$\pm$0.07 & 0.77$^{+0.34}_{-0.21}$ & 0.63 -- 1.00 & 1.69$^{+0.48}_{-0.34}$ \\
809741149368202752 & 1.38$\pm$0.08 & 14.91 & 922$\pm$101 & 0.35$\pm$0.07 & 0.77$^{+0.17}_{-0.13}$ & 0.91$\pm$0.2 & 1.76$^{+0.27}_{-0.24}$ \\
5847919241396757888 & 1.46$\pm$0.15 & 16.90 & 1254$\pm$290 & 0.68$\pm$0.11 & 0.86$^{+0.88}_{-0.39}$ & 0.63 -- 1.00 & 1.80$^{+1.05}_{-0.56}$ \\
5580526947012630912 & 1.13$\pm$0.02 & 13.36 & 654$\pm$10 & 0.76$\pm$0.08 & 0.68$^{+0.37}_{-0.21}$ & 1.13$\pm$0.2 & 1.80$^{+0.55}_{-0.35}$ \\
1350295047363872512 & 1.12$\pm$0.03 & 13.52 & 657$\pm$9 & 0.66$\pm$0.07 & 0.57$^{+0.11}_{-0.09}$ & 1.44$\pm$0.2 & 1.83$^{+0.22}_{-0.19}$ \\
4744087975990080896 & 1.99$\pm$0.12 & 17.07 & 631$\pm$11 & 0.61$\pm$0.09 & 0.87$^{+0.35}_{-0.25}$ & 0.63 -- 1.00 & 1.83$^{+0.48}_{-0.39}$ \\
6001459821083925120 & 0.89$\pm$0.04 & 13.60 & 564$\pm$13 & 0.47$\pm$0.10 & 0.86$^{+0.45}_{-0.29}$ & 0.63 -- 1.00 & 1.81$^{+0.58}_{-0.44}$ \\
1749013354127453696 & 0.61$\pm$0.05 & 14.49 & 932$\pm$155 & 0.51$\pm$0.15 & 0.84$^{+0.64}_{-0.42}$ & 1.00$\pm$0.2 & 1.93$^{+0.86}_{-0.63}$ \\
4314242838679237120 & 2.00$\pm$0.22 & 17.02 & 1146$\pm$382 & 0.70$\pm$0.09 & 1.22$^{+1.65}_{-0.64}$ & 0.63 -- 1.00 & 2.25$^{+1.87}_{-0.84}$ \\
5593444799901901696 & 0.60$\pm$0.05 & 14.42 & 1039$\pm$292 & 0.44$\pm$0.14 & 1.15$^{+0.66}_{-0.47}$ & 1.27$\pm$0.2 & 2.57$^{+0.86}_{-0.69}$ \\
6328149636482597888 & 1.23$\pm$0.08 & 13.34 & 736$\pm$23 & 0.14$\pm$0.07 & 1.29$^{+1.28}_{-0.23}$ & 1.21$\pm$0.2 & 2.71$^{+1.50}_{-0.36}$ \\
\enddata
\tablenotetext{a}{Quoted uncertainties are 2$\sigma$ or 95\% confidence intervals.}
\tablenotetext{b}{For primary stars with masses derived by either our UCO Lick spectra or from Apsis, we adopt a mass uncertainty of $0.1$~\Msun. For any remaining stars without mass measurements, we assume primary masses lie somewhere between 0.63 and 1~\Msun. }
\end{deluxetable*}

We further refine this sample of 106 candidates by adding an additional set of quality constraints. 
First, we only select binaries with a goodness-of-fit ($F_2$) less than five. We further add the restriction that $M_2/\sigma_{M_2} > 3$, to select only binaries with relatively well-measured companion masses. Finally, we add two separate criteria to remove any giant star donors. This constraint may seem overly constrictive since giant stars with NS and BH companions ought to exist in the Milky Way \citep{breivik2019}, and indeed the only known BH binary in the field hosts a giant star \citep{thompson2019}. Nevertheless, the large luminosities of giant stars can hide even potentially massive, non-degenerate companions \citep{elbadry2022b}. We therefore select for only dwarf stars by first removing any stars with a log $g$ $<$ 3.6 \citep[if the stars have a log $g$ measured by Apsis, the Gaia team's astrophysical parameters inference system;][]{GaiaDR3_apsis, GaiaDR3_apsis2} and second making a cut in the color-magnitude diagram:
\begin{equation}
G > 3.14 (BP-RP) - 0.43,
\end{equation}
where $G$, $BP$, and $RP$ are absolute magnitudes (we have not accounted for dust extinction), derived from each star's apparent magnitudes and parallax. 

The top panel of Figure~\ref{fig:dwarfs} shows the $T_{\rm eff}$ and log $g$ values for the stars which have been measured by Apsis, while the bottom panel shows a color-magnitude diagram with PARSEC isochrones overplotted \citep{bressan2012}. Note that we have not taken into account reddening and extinction in this sample which can be significant as many of these stars lie near the Galactic Plane.

\begin{deluxetable*}{lcccccc}
 \tablecaption{ Log of spectroscopic observations taken on 2022-06-24 \label{tab:kastlog}}
 
 \tablehead{
    \colhead{Gaia DR3 Source ID} & 
    \colhead{R.A.} & 
    \colhead{Dec.} & 
    \colhead{Start Time} &
    \colhead{Airmass} & 
    \colhead{Exposure} & 
    \colhead{Slit} \\
    \colhead{} &
    \colhead{} &
    \colhead{} &
    \colhead{(UT)} &
    \colhead{} &
    \colhead{(s)\tablenotemark{a}} &
    \colhead{P.A.} 
    }
\startdata
3649963989549165440 & 14:33:30.86 & $-$01:14:42.97 & 06:23 & 1.45 & 630, 2$\times$300 & 32  \\
1581117310088807552 & 12:20:12.81 & $+$58:41:16.44 & 05:00 & 1.19 & 630, 2$\times$300 & 111 \\
1947292821452944896 & 21:46:56.56 & $+$33:28:13.87 & 11:05 & 1.01 & 730, 2$\times$350 & 118 \\
1350295047363872512 & 17:39:56.06 & $+$45:02:17.33 & 10:51 & 1.25 & 530, 2$\times$250 & 85  \\
1525829295599805184 & 13:13:29.01 & $+$41:51:54.00 & 05:34 & 1.11 & 930, 2$\times$450 & 87  \\
4373465352415301632 & 17:28:41.08 & $-$00:34:51.93 & 08:08 & 1.29 & 630, 2$\times$300 & 194 \\
1854241667792418304 & 21:27:37.70 & $+$33:16:23.86 & 11:25 & 1.00 & 630, 2$\times$300 & 163 \\
1749013354127453696 & 20:33:15.26 & $+$07:58:46.02 & 11:43 & 1.19 & 530, 2$\times$250 & 205 \\
\enddata
\tablenotetext{a}{Exposure sequence. The comma denotes separate sequences for the blue and red channels of Kast, respectively.}
\end{deluxetable*}

Within our resulting sample, we remove one particular system, Gaia DR3 4373465352415301632, as it has a large $m_f$ of $\simeq$11.6\Msun\ and an orbital period of 186 days, or roughly three times Gaia's scanning period. \citet{halbwachs2022} have identified systems with large $m_f$ and orbital periods that are harmonics of Gaia's scanning law as being possible contaminants. After removing this system, our sample contains 24 candidate compact object binaries, which we provide in Table~\ref{tab:sample}.

\section{Observations}
\label{sec:observations}

As a test of the validity of our sample, we obtained spectroscopic follow-up observations of eight of our sample of 24 candidate compact object binaries using the Kast spectrograph \citep{Miller93} mounted on the Shane 3-m telescope on 2022 Jun 24 (UT). We used the 300/7500 grating on the red side and the 452/3306 grism on the blue side, providing continuous coverage between 3500--10750 \AA. We used a 2\arcsec\ slit to match the seeing conditions, which were 1.4\arcsec\ at the beginning of the night, but was variable throughout the night. The slit was orientated to the parallactic angle for all exposures to avoid differential flux losses. A log of our observations is presented in Table \ref{tab:kastlog}.

Kast spectra were reduced in a standard manner using the \textsc{IRAF} routines within a custom pipeline\footnote{\url{https://github.com/msiebert1/UCSC_spectral_pipeline}}, including bias subtraction, flat fielding, wavelength calibration, cosmic-ray rejection and spectral extraction. Spectroscopic flux calibration was performed relative to standard stars at a similar airmass to the targets and using the same observational setup. BD$+$284211 and BD$+$262606 were used to flux-calibrate and telluric-correct the blue and red spectra, respectively.  Using the fluxes within the $\sim$150~\AA\ overlap region between the two sides, we rescaled, interpolated, and combined the blue and red portions of the Kast spectra to produce a single flux-calibrated spectrum for each star.  More details for these procedures are provided in \citet{Silverman12}.

\begin{figure*}
    \begin{center}
    \includegraphics[width=0.9\textwidth]{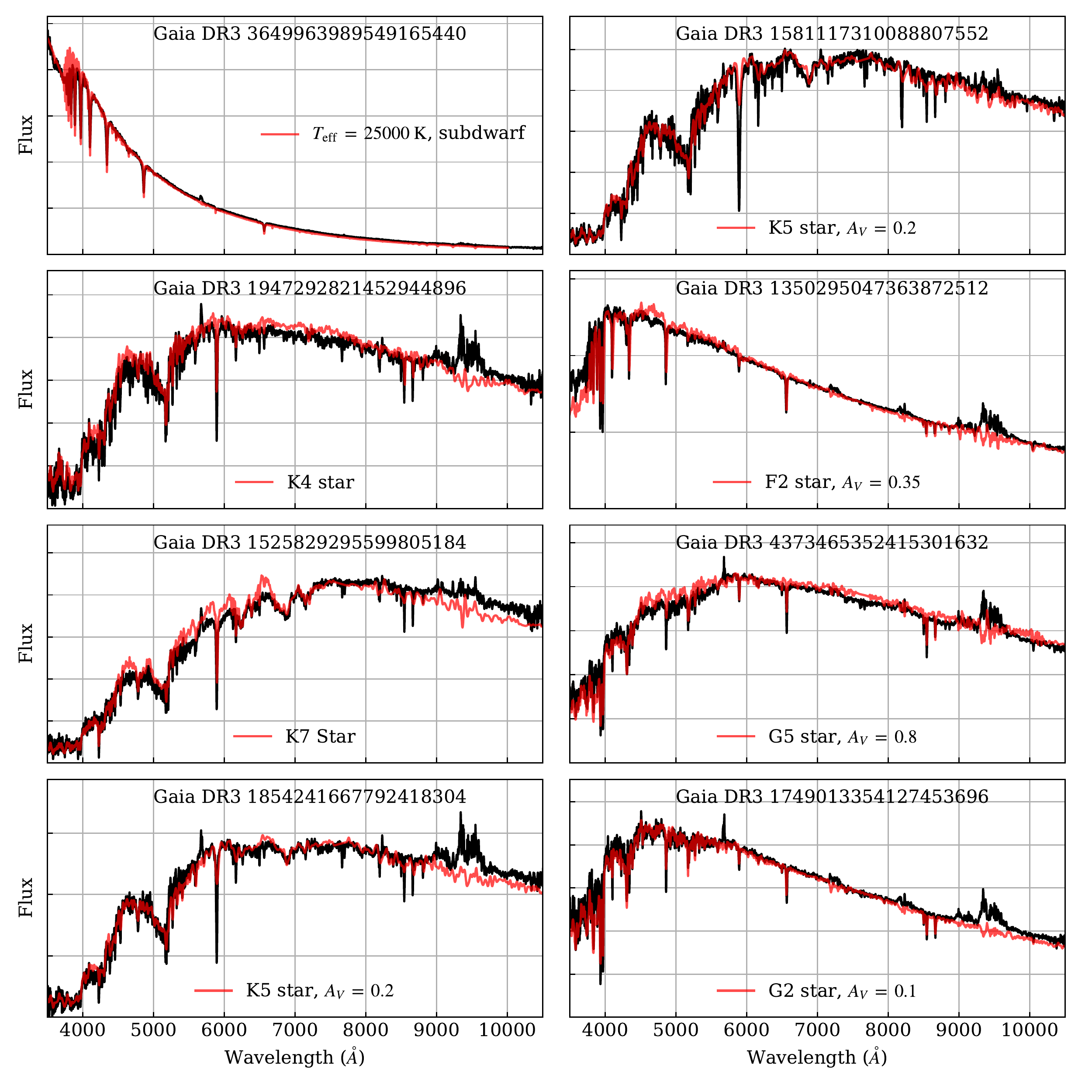}
    \caption{ Optical spectra of eight stars from our list of 24 candidates compared with templates from \citet{pacheco2021} and \citet{pickles1998}. The top left panel shows that Gaia DR3 3649963989549165440 is consistent with a being a hot subdwarf, originally classified as such by \citet{boudreaux2017} and \citet{geier2017}. The other seven stars are consistent with main sequence dwarf stars, none of which show significant excess emission at blue wavelengths, indicating that there is no evidence for the existence of a hidden white dwarf in the system. Note that on the red side, there is a ghost image caused by reflections in the optics that overlaps with the primary spectrum, causing inaccurate flux calibration over 5630–5730~\AA. While we have attempted to mitigate this issue with careful treatment of the spectra in this area, residuals remain and the flux measurements at these wavelengths, and we have ignored these data in our analysis.}
    \label{fig:spectra}
    \end{center}
\end{figure*}

In Figure~\ref{fig:spectra} we provide the reduced spectra for these eight stars along with spectral templates for comparison for the hot subdwarf from \citet{pacheco2021} and for the dwarfs from \citet{pickles1998}. Where indicated in the figure, we have added dust extinction to the spectral templates using the python package {\tt extinction} \citep{extinction} which applies the extinction curve from \citet{fitzpatrick2007}. While most stars are F-, G-, and K-dwarfs, Gaia DR3 3649963989549165440 is a hot star, consistent with being a subdwarf (we discuss this system further in Section~\ref{sec:discussion}). Importantly, none of these spectra show excess flux at short wavelengths, which would be indicative of the presence of a hidden white dwarf. Furthermore, all appear to be well-fit by a single spectral component; there is no obvious evidence of a second luminous star contributing to the observed flux.

The spectral templates provided in each panel of Figure~\ref{fig:spectra} are not fits, but rather by-eye approximations to the observations. We make the reduced spectra freely available for users who would like to perform their own fits to models. We use these spectral fits to update the primary masses of the systems in our sample, which we then propagate to derive updated compact object masses.

\begin{figure}
    \begin{center}
    \includegraphics[width=1.0\columnwidth]{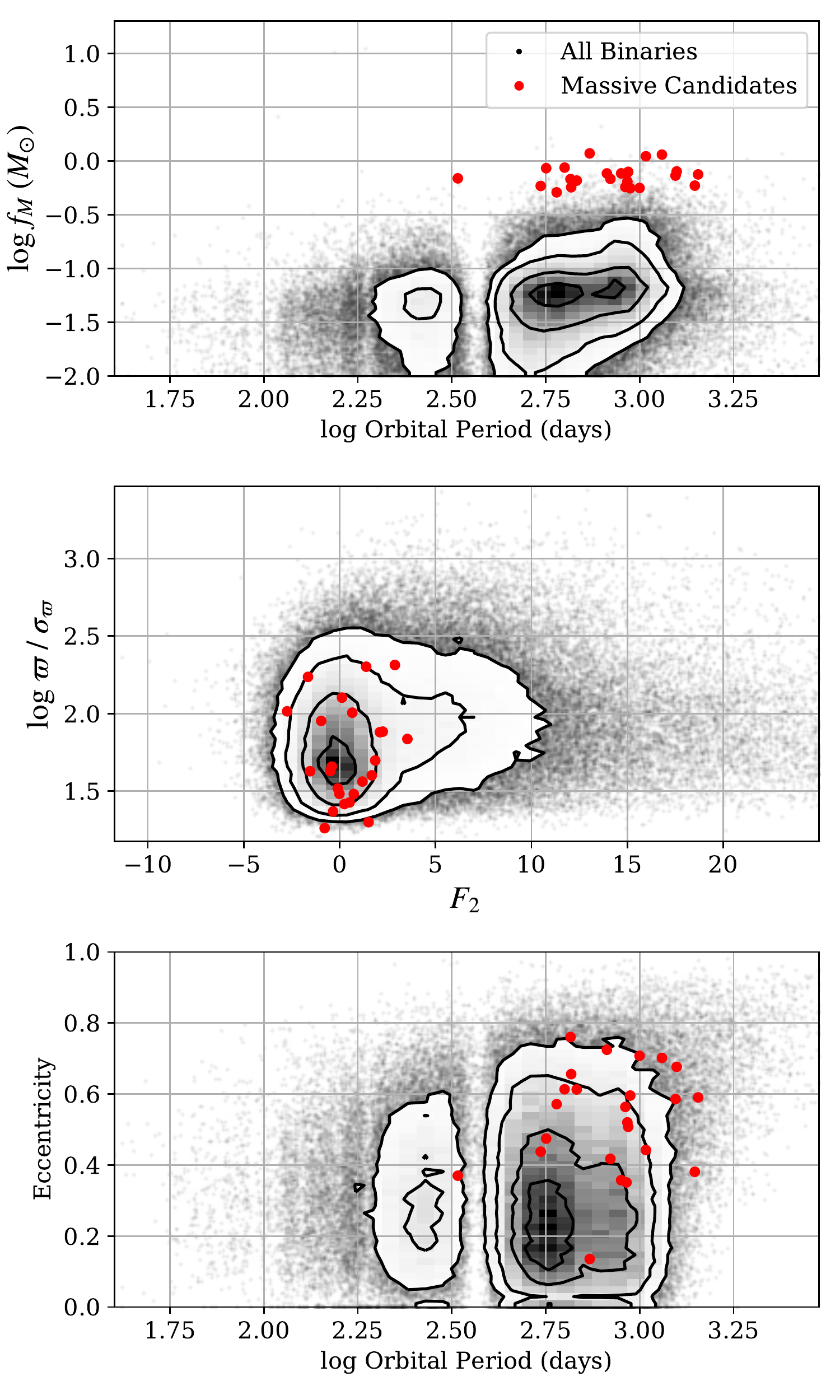}
    \caption{ We compare the mass function with the orbital period of the entire catalog of astrometric binaries in Gaia (black contours and points) and our sample of candidate massive binaries (red) in the top panel. The middle panel compares the goodness-of-fit with the parallax significance. Our selection criteria use include a conservative constraint that $F_2<5$, even though the Gaia catalog extends to $F_2$ of 25. The bottom panel shows that the binaries in our sample tend to have orbital periods of a few years and eccentricities above 0.4.}
    \label{fig:sample_summary}
    \end{center}
\end{figure}

\section{Discussion and Caveats}
\label{sec:discussion}

\begin{figure}
    \begin{center}
    \includegraphics[width=1.0\columnwidth]{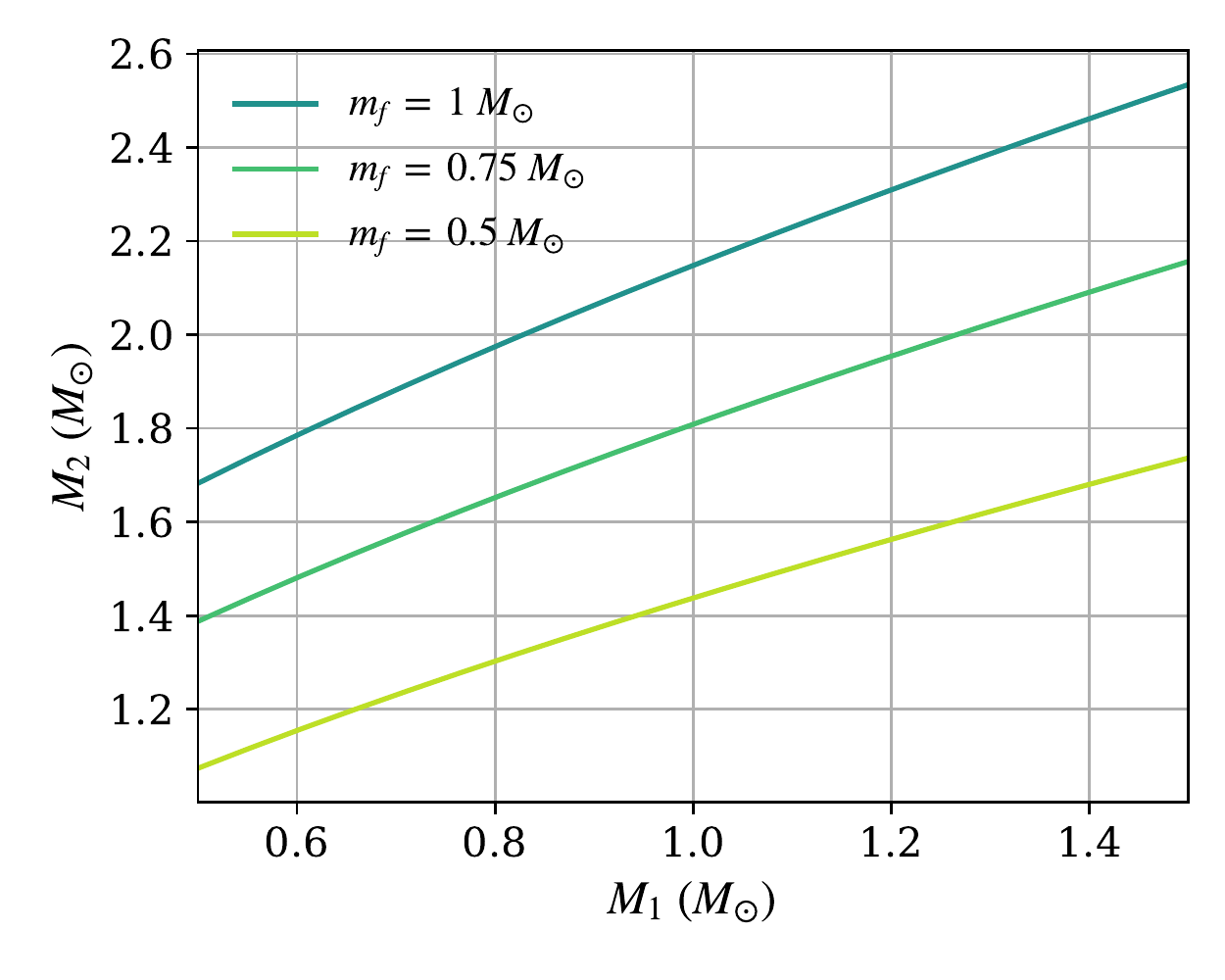}
    \caption{ Distribution of compact object masses as a function for luminous star masses for three different mass functions. The mass functions in our sample range from 0.51 to 1.28~\Msun (see Table~\ref{tab:sample}). For those stars in our sample where we lack mass measurements for the luminous star, we assume a mass of 1~\Msun; however, even if these stars are late K-dwarfs with masses of 0.7~\Msun, they are still likely to host NS companions with masses $\gtrsim$1.2 \Msun. }
    \label{fig:mf}
    \end{center}
\end{figure}

Figure~\ref{fig:sample_summary} shows various summary statistics for our sample of 24 candidate compact object binaries. In the top panel we show $m_f$ as a function of orbital period, while the middle panel shows the parallax significance ($\varpi/\sigma_{\varpi}$) as a function of $F_2$. While the Gaia data contain orbital solutions for binaries with $F_2<25$, we find many of the solutions with $F_2$ down to at least 10 demonstrate spurious orbits. To reduce the possibility of sample contamination, we restrict our catalog so our binaries all have $F_2 < 5$. Compared with the background sample showing all astrometric binaries in the Gaia catalog, our quality cuts are quite conservative. Finally, in the bottom panel of Figure~\ref{fig:sample_summary} we provide the orbital periods and eccentricities of our sample. The binaries in our catalog tend to be on very long orbits (periods of years) with eccentricities typically above 0.4. 

Our classification that these binaries host compact objects is dependent first and foremost on the companion masses we derive, which itself depends on the primary masses we adopt. For the subset of stars with masses provided by Apsis and for those which we have our own spectroscopic follow-up observations, we expect our mass estimates are accurate to within $\lesssim$0.1~\Msun. The three stars with masses measured by both methods exhibit agreement to with a root-mean square of $\sim$0.1~\Msun. When deriving companion masses, we therefore adopt an uncertainty of 0.1~\Msun\ for the masses of the luminous stars. For the remainder of our sample, we adopt a luminous star's mass from a uniform distribution ranging from 0.63 to 1.00~\Msun. This is an obvious inaccuracy that we plan to address with future spectroscopic follow-up. 

In Figure~\ref{fig:mf} we show the dependence of the derived $M_2$ on $M_1$ for three different mass functions. Variations of 0.1~\Msun\ in $M_1$ lead to similar differences of $\simeq$0.1~\Msun\ in $M_2$, suggesting our sample is robust to these inaccuracies in $M_1$, at least for the subset of our sample where it is measured. Nevertheless, Figure~\ref{fig:mf} provides some idea of how inaccurate our derived $M_2$ measurements could be. For a system with a $m_f$ of 0.5~\Msun, assuming a G2 star with a mass of 1~\Msun\ would imply a companion mass of 1.44~\Msun. If that star is actually a mid K-dwarf with a mass of 0.7~\Msun, the implied companion mass is 1.23~\Msun. Although WDs in excess of 1.3~\Msun\ probably exist in stellar binaries \citep[e.g., SDSS J0811$+$0225;][]{brown2013}, the WD mass distribution implies they are extremely rare \citep{tremblay2016}, and our interpretation that a particular system is likely to host a NS or BH is still valid.

\begin{figure}
    \begin{center}
    \includegraphics[width=1.0\columnwidth]{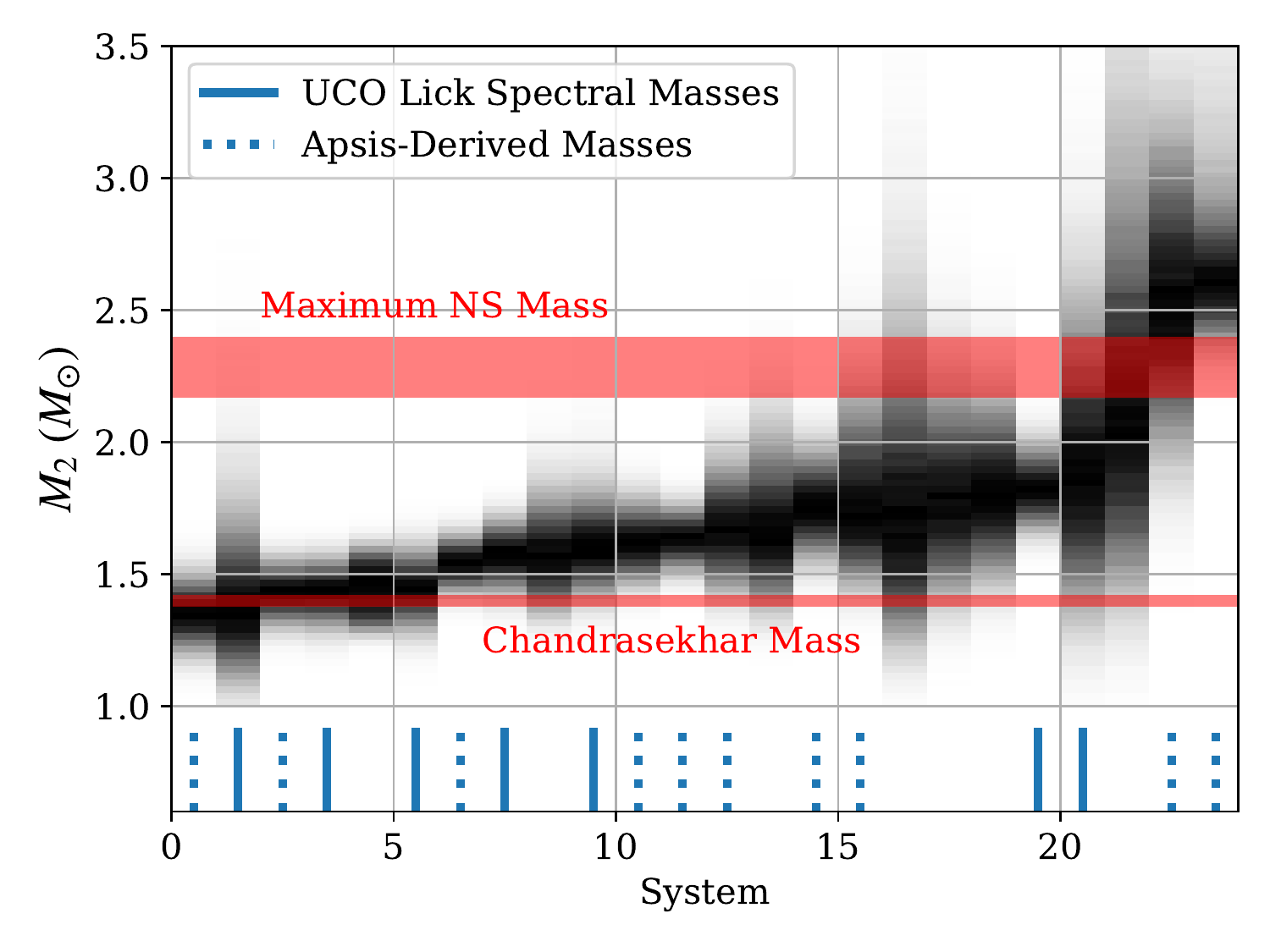}
    \caption{ Distribution of companion mass likelihoods for each of the 24 candidate binaries in our sample. Each column consists of one binary from our sample and the grayscale color denotes the mass probability. Tick marks on the $x$-axis indicate whether we used a mass of the primary derived from the spectra shown in Figure~\ref{fig:spectra} (solid ticks) or from Apsis (dotted ticks). Systems in which neither measurement are available have no tick marks, in which case the companion mass is derived from assuming a uniform distribution of possible primary masses between a range of 0.6 to 1~\Msun. Our sample of companion masses span the range of 1.35 to 2.7~\Msun. }
    \label{fig:m2}
    \end{center}
\end{figure}

Using the posterior samples from our derived mass functions and our best estimates for the luminous star's mass, following the procedure in Section~\ref{sec:sample_selection}, but this time propagating uncertainties in the luminous stars' measured masses, we derive the distribution of companion masses. We provide the median derived compact object masses and their 95\% confidence intervals as the last column in Table~\ref{tab:sample}. We graphically display the distribution of companion masses in Figure~\ref{fig:m2}; each column represents one system, with the colorscale representing the probability distribution of each system's companion mass. We use tick marks on the $x$-axis of this figure to indicate whether the masses of the luminous stars were derived from our follow-up spectra (solid ticks), Apsis (dotted ticks), or our minimally assumptive mass range (no ticks). The sample spans a mass range of 1.35 to 2.7~$\Msun$ and therefore likely contains a combination of both NSs and BHs. For reference, we show both the Chandrasekhar mass and an estimated range for the maximum NS mass of 2.17~\Msun\ \citep[as derived from the electromagnetic counterpart to the gravitational wave event GW170817;][]{margalit2017} to 2.4~\Msun\ \citep[predicted by certain equations of state;][]{ozel2016}.

What are the possible contaminants in our sample? Our analysis relies heavily on the astrometric fits provided by Gaia. However, in constructing our sample, we find there are many problematic binaries within the Gaia astrometric binary catalog itself. As previously mentioned, Gaia DR3 4373465352415301632 provides one example. Despite its relatively high significance and low $F_2$ score, we removed this binary from our sample by hand as it has an orbital period close to three times Gaia's scanning law. Gaia DR3 3640889032890567040 and Gaia DR3 3545469496823737856 provide additional examples. Taken at face value, the astrometry of these systems implies they have companions with masses in excess of $\simeq$100~\Msun. However, Gaia DR3 3545469496823737856 has a derived parallax of 76 mas, implying it is within the nearest 15 pc and Gaia DR3 3640889032890567040 has a goodness-of-fit of 10.3 as well as abnormally large errors on the measured $F$ and $G$. Furthermore, these particular solutions have singular covariance matrices. We therefore consider these systems' orbital fits to be unreliable. Because the astrometric binary catalog only provides fits to the data without either the posterior samples or the time series astrometric data from which the fits were made, determining the validity of individual astrometric binaries is challenging. While we cannot guarantee our sample is free from contamination due to bad orbital fits, we have made every effort to do so, choosing conservative quality cuts at every step. 

We have also made every effort to ensure that the binaries in our sample host NS and BH companions. WDs are possible contaminants, however they would need to be extremely heavy (relative to the WD mass distribution). Furthermore for the eight systems (a third of our sample) where we have obtained follow-up spectroscopy, there is no evidence for excess emission at short wavelengths. We have also checked the BP and RP spectra provided by Gaia \citep{GaiaDR3_BPRP} for those stars in our sample where they are available and none show significant excess at blue wavelengths. Finally, WD companions are unlikely on theoretical grounds. Depending on their masses, WDs in binaries at these orbital periods should have gone through Roche lobe overflow prior to becoming a WD. Binary evolution codes commonly assume that a binary circularizes due to mass transfer and tidal forces as soon as it overfills its Roche lobe \citep[e.g.,][]{hurley2002}. The bottom panel of Figure~\ref{fig:sample_summary} shows these systems are all eccentric. Combined, these pieces of evidence implies that WDs provide, at worst, a small source of contamination.

The companions in each of our binaries cannot be main sequence stars, as these would all be significantly luminous --- in every case more luminous than the observed star. It is also unlikely that the companion is an evolved star, which would produce substantial flux, or a massive stripped star, which would produce a blue excess in our spectra.  None of the spectra in Figure~\ref{fig:spectra} show any evidence for emission from a second component. Future analysis using more sophisticated fitting techniques are required to derive statistical limits. 

We finally consider the possibility that the companion is actually itself a tight binary of two G- or K-dwarfs. We also find this possibility unlikely for three reasons: 1) we would expect the stars to be over-luminous in the color-magnitude diagram, but Figure~\ref{fig:dwarfs} shows most of the stars in our sample lie close to the expected Main Sequence for single stars. 2) Two (or more) luminous companion stars means Gaia observes the motion of the photocenter, not the primary's motion, implying a larger orbit--and therefore even more massive companion system--based on the results shown in Figure~\ref{fig:mass_function}. 3) Such hierarchical triples are only stable only for a limited time; with an outer binary separation of $\sim$AU, the inner orbit would have to be $\sim$\Rsun\ to satisfy the stability condition from \citep{naoz2014}. 

We therefore conclude that contamination in our sample by non-NS and non-BH companions is likely to be minimal.

The system Gaia DR3 3649963989549165440 is worthy of particular attention as it hosts a hot subdwarf system with a putative NS companion. This object is contained in previous catalogs of subdwarfs \citep{boudreaux2017, geier2017, geier2019}, and our follow-up spectrum shown in the top left panel of Figure~\ref{fig:spectra} is consistent with that classification. In deriving its companion mass of 1.41~\Msun, we have assumed a canonical mass of 0.47~\Msun\ for the subdwarf \citep{zhang2009, heber2016}. A more careful spectral analysis will need to ensure that a 1.4~\Msun\ F-star cannot be hidden by the luminous subdwarf. If more detailed spectra rule out that possibility, then this system forms a rare subdwarf/NS binary with no previously known examples. Since subdwarfs may be the result of stripping during binary evolution \citep{heber2016}, this binary forms a unique test bed for binary evolution studies.

The objects in our catalog span a range of masses, from 1.35 to 2.7~\Msun. The most massive objects in our sample are very likely above the maximum NS mass; our sample therefore fills in part of the so-called lower mass gap, demonstrating a smooth mass transition from NSs to BHs. Furthermore, the existence of $\simeq$1.7~\Msun\ NSs in the Milky Way is also interesting. NSs of 2~\Msun\ are known to exist \citep{demorest2010, antoniadis2013, cromartie2020}, but these are all found in close binaries around low-mass stars, where the NS is expected to have grown substantially through accretion. The NSs and BHs in our sample are in binaries too wide to have grown their mass through accretion, and were therefore born with their current masses. The mass of the NS in Vela X-1, which is likely too young to have accreted significant material, is similarly massive \citep{barziv2001, quaintrell2003}. However, none of the NSs in the $\simeq$20 known double NSs, are more massive than $\simeq$1.56~\Msun\ \citep{tauris2017}. The existence of an additional population of $\simeq$1.7~\Msun\ NSs may help understand the gravitational wave source GW190425, which was comprised of two compact objects with a total mass of 3.4~\Msun\ \citep{GW190425}. While the system could have been a NSBH system with masses of 1.4 and 2.0~\Msun\ \citep{foley2020}, the presence of 1.7~\Msun\ NSs in our sample suggest that a double NS scenario for GW190425 is possible.

Finally, we comment that our sample is biased toward higher compact object masses ($>1.4~\Msun$), as we are purposefully searching for binaries that are inconsistent with hosting massive WD companions. We plan to expand this analysis in a future study to lower mass companions to identify a more complete sample of NS binaries. At the same time, higher-mass BHs ought to be easier to identify as they produce larger orbits, the key criterion for astrometric detectability \citep{andrews2019}. Although we refrain from making detailed population conclusions due to the complex observational biases involved, the lack of higher-mass BHs strongly suggests that binaries hosting these objects are quite rare in the Solar Neighborhood.

\section{Conclusions}
\label{sec:conclusions}

We have analyzed the catalog of astrometric binaries in Gaia DR3 to search for systems likely to host NS or BH companions. We have purposefully made conservative quality constraints to reduce contamination from either bad astrometric solutions or non-compact object binaries. Our catalog contains 24 systems with companions with masses ranging from 1.35 to 2.7~\Msun. To aid in ruling out the possibility of contamination from non-NS and non-BH companions, we have taken follow-up spectra of eight of the stars in (a third of) our sample, finding no evidence for the presence of any additional components. We consider several possible types of contaminating companions, but none are fully consistent with the observations.

Because the procedure for producing our catalog is comparatively stringent, we are likely incomplete, and additional NS- and BH-hosting binaries probably exist within the Gaia dataset.  Furthermore, the Gaia dataset itself has additional systematic biases that are difficult to fully understand without the full astrometric data set, instead of the currently released best-fit parameters. Therefore, one should take care when deriving population results using this sample. Nevertheless, there are a few conclusions that can be made from the existence of even some binaries with the observed characteristics. 

For instance, the orbital velocities of the binaries in this sample are $\sim$20 km s$^{-1}$ or roughly an order of magnitude lower than the typical velocities that NSs receive at birth \citep{hobbs2005}. Had these NSs been formed with kicks of 265 km s$^{-1}$, most would have disrupted, and any survivors would be characterized by very high eccentricities \citep{andrews2019b}. It is well known that some NSs are formed with low kick velocities \citep{wong2010}, but previous mechanisms involve either an electron-capture \citep{nomoto1984} or ultra-stripped SN \citep{tauris2015}, both of which are expected to form low-mass NSs. The NSs in this sample are comparatively massive. We therefore conclude that at least some massive NSs ($M\gtrsim$1.5~\Msun) are formed with low kick velocities. 

It is also worth considering the future evolution of these binaries. Even the most massive primary stars in our sample will take Gyr to evolve off the Main Sequence, and when they do so they will evolve into white dwarfs. Given their low masses and wide orbits, these binaries will not become gravitational wave sources within a Hubble time unless an additional phase of mass transfer can bring the orbits into a much closer configuration. We leave a more detailed analysis of the evolution of these binaries for a future work, and for now we only comment that our sample implies the existence of white dwarfs with bound BH companions. Such systems could be identified in future astrometric or radial velocity catalogs.

The objects in this sample are ripe for observational follow-up across the electromagnetic spectrum. At long wavelengths, these binaries form ideal targets for radio searches for pulsed emission from our NS-mass companions. A more complete ultraviolet, optical, and infrared follow-up of our sample can aid in ruling out exotic scenarios involving non-NS and non-BH companions. Finally, at the shortest wavelengths, X-ray observations can be used to place constraints on accretion models in the radiatively inefficient regime.

Future Gaia data releases will contain even larger samples of astrometric binaries, and will likely only expand the sample of dark compact objects in wide orbits around luminous stars. In the meantime to aid users' ability to consider the validity of individual systems' orbital solutions, we urge the Gaia team to release time series astrometric data for these systems.

\acknowledgements

We thank Grace Yang for assisting with the Shane observations.

J.J.A.\ acknowledges support from CIERA and Northwestern University through a Postdoctoral Fellowship. This work was performed in part at the Aspen Center for Physics, which is supported by National Science Foundation grant PHY-1607611.  J.J.A.\ and R.J.F.\ initiated their collaboration during the ``Fundamental Physics and Astrophysics with the Next Generation of Gravitational-Wave Detectors'' program, of which they were both participants.

The UCSC team is supported in part by NSF grant AST--1815935, the Gordon \& Betty Moore Foundation, the Heising-Simons Foundation, and by a fellowship from the David and Lucile Packard Foundation to R.J.F.

We would like to express our gratitude to the Lick Observatory staff for their support, in particular Matt Brooks and Elinor Gates. We gratefully acknowledge usage of native lands for our science. Shane 3-m observations were conducted on the stolen land of the Ohlone (Costanoans), Tamyen and Muwekma Ohlone tribes. 

A major upgrade of the Kast spectrograph on the Shane 3-m telescope at Lick Observatory was made possible through generous gifts from William and Marina Kast as well as the Heising-Simons Foundation. Research at Lick Observatory is partially supported by a generous gift from Google.

\software{{\tt astropy} \citep{astropy}, {\tt corner} \citep{corner}, {\tt extinction} \citep{extinction}, {\tt NumPy} \citep{numpy}, {\tt SciPy} \citep{scipy}, {\tt matplotlib} \citep{matplotlib}}

\bibliographystyle{aasjournal}
\bibliography{gaia}

\end{document}